# Pulsed field studies of the magnetization reversal in molecular nanomagnets

J. Vanacken[1], S. Stroobants[1], M. Malfait[1], V.V. Moshchalkov[1],
[1]*Pulsed field group, Laboratory of Solid-State Physics and Magnetism, K.U.Leuven , Celestijnenlaan 200D, B-3001 Heverlee, Belgium*

M. Jordi[2], J. Tejada[2], R. Amigo[2],
[2]*Department de Física Fonamental, Universitat de Barcelona, Diagonal 647, 08024 Barcelona, Spain*

E. M. Chudnovsky[3] and D.A. Garanin[3,4]
[3]*Department of Physics, Lehman College – CUNY, Bronx, NY 10468-1589, U.S.A.*
[4]*Institut für Physik, Johannes-Gutenberg-Universität, D-55099 Mainz, Germany*

We report experimental studies of crystals of $Mn_{12}$ molecular magnetic clusters in pulsed magnetic fields with sweep rates up to $4 \times 10^3$ T/s. The steps in the magnetization curve are observed at fields that are shifted with respect to the resonant field values. The shift systematically increases as the rate of the field sweep goes up. These data are consistent with the theory of the collective dipolar relaxation in molecular magnets.

High-spin molecular nanomagnets, like $Mn_{12}$ acetate, have unusual magnetic properties related to their high magnetic anisotropy and to the quantization of the magnetic moment **M**. For certain values of the magnetic field, quantum states characterized by different projections of **M** onto the anisotropy axis come to resonance. At these fields the magnetization curve of the crystal exhibits distinct steps due to quantum transitions between the resonant energy levels [1]. The steps, for a field-sweep experiment, have been successfully described in terms of single-molecule Landau-Zener (LZ) transitions [2-7]. To date the information about spin Hamiltonians, extracted from the magnetization measurements [1,8-11], has been compared with the EPR data [12-17] and a good agreement has been achieved.

In this Letter we report low temperature magnetization studies of $Mn_{12}$ single crystals at a field sweep rate $\mu_0\, dH/dt$ up to 4 kT/s. Our main finding is that at such high sweep rates the position of the steps in the magnetic relaxation shifts by ΔH that increases as the sweep rate goes up. We have been able to scale the relaxation curves obtained at different sweep rates onto one curve. The scaling can be explained within a model of collective magnetic relaxation of the crystal, suggested in [18].

$Mn_{12}$ single crystals of high purity were used in the experiments. The conventional composition and the structure of the crystals were established by chemical, infrared and X-ray diffraction methods. In addition, dc and ac magnetometry of the crystals was carried out in order to verify their conventional behaviour at low sweep rates. We have checked that the values of the blocking temperatures and resonant fields of the crystals coincide with previously published values.

Measurements of the magnetization using fast magnetic field pulses up to 4 kT/s and at a temperature $T$=0.6 K were performed at the K.U.Leuven. The pulsed magnetic fields were generated by a modular capacitor bank whose capacitance was systematically tuned from $C$=4 mF to $C$=28 mF while the voltage was adapted from $V$=5000 V to $V$=600 V in such a way that the capacitor energy, $\frac{1}{2}CV^2$, remained constant. A home-made coil with an inductance of 650 μH was used to produce the magnetic field pulse. A crow bar diode of resistance $R$ = 0.08 Ω provided a critical damping of the magnetic pulse which has a duration of ~ 20 ms. The magnetization measurements were performed with the use of an inductive magnetization sensor designed to measure samples of volume up to 1 mm$^3$. The sensor coil had 640 turns in one direction and 345 turns in the opposite direction. The sensitivity of this probe reaches $10^{-4}$ emu in the fields up to 10 T. During the measurements, the sample and the detection coils were submerged in liquid $^3$He. The $^3$He temperature probe was made entirely of non-metallic materials; we have verified that during a 50 T field pulse the temperature change, measured by a calibrated RuO sensor, did not exceed 100 mK. The sweep rate versus magnetic field of a typical magnetic field pulse is shown in Figure 1.

The typical field dependence of the differential susceptibility, $dM/dH$, of a single crystal of $Mn_{12}$ acetate, taken at various sweep rates and $T$ = 670 mK, is shown in Figure 2. The magnetization reversal occurs at a field that is close to the third resonant field, $\mu_0\, H$ ~ 1.3 T [1]. The most surprising feature of the data is the dependence of the position and the height of the peaks on the field sweep rate. According to the conventional theory of resonant spin tunneling [1], confirmed by all previous experimental studies, the positions of the peaks are determined entirely by the Hamiltonian of the nanomagnet and should not depend on the sweep rate. Note that some dependence of the peaks on the rate may occur in the case of thermal avalanches [19-22]. In this case, however, the magnetization reversal is always accompanied by the measurable increase of the temperature of the sample. In our experiments no significant change in the temperature has been detected, making avalanches an improbable explanation. As an additional argument disarming the heating scenario, one should notice that when heating occurs, tunneling will occur from levels further away from the ground state; which would mean that peaks should shift to the left (lower fields) with higher dH/dt (supposing increased heating) in stead of the observed shift to the right (higher fields).

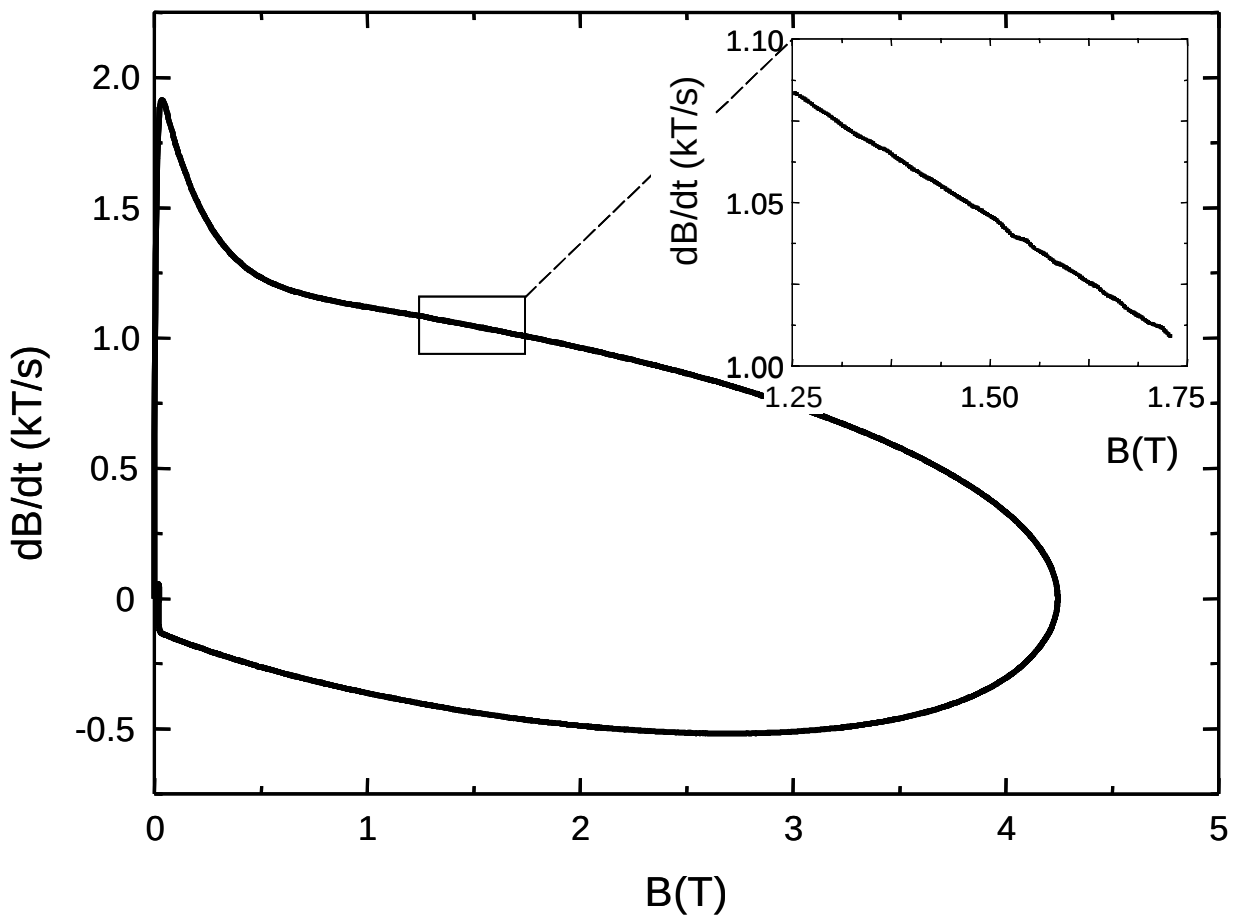

Figure 1. Field dependence of the sweep rate.

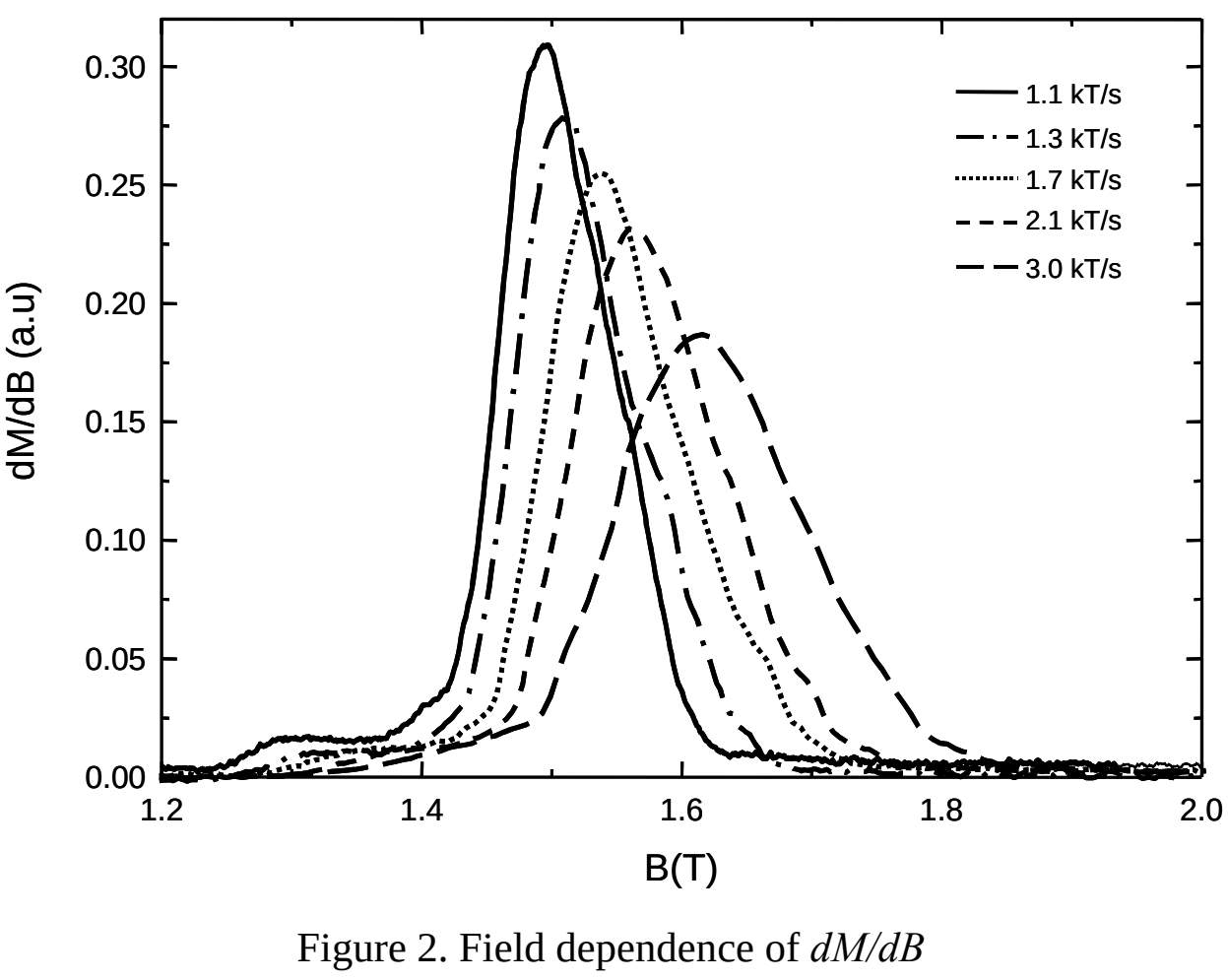

Figure 2. Field dependence of *dM/dB* for different sweep rates at B=1.3T.

One possible explanation to the above findings can be obtained along the lines of the collective magnetization reversal expected at a very high sweep rate. According to [22], for such a sweep the relaxation of the magnetization at the level crossing occurs in two stages. The first stage is the Landau-Zener process that leaves the fraction of magnetic molecules $P$ in the excited states (the upper energy branch $\varepsilon^+$ in Fig. 3). This fraction is given by the Landau-Zener formula: $P_{LZ}=exp(-\varepsilon)$, where $\varepsilon= \pi\Delta^2/2\hbar v$, $\Delta$ is the tunnel splitting and $v$ is the energy sweep rate $W=vt= g\ |\Delta m|\ \mu_B\ (H(t)-H_R)=\varepsilon_m-\varepsilon_{m'}$.

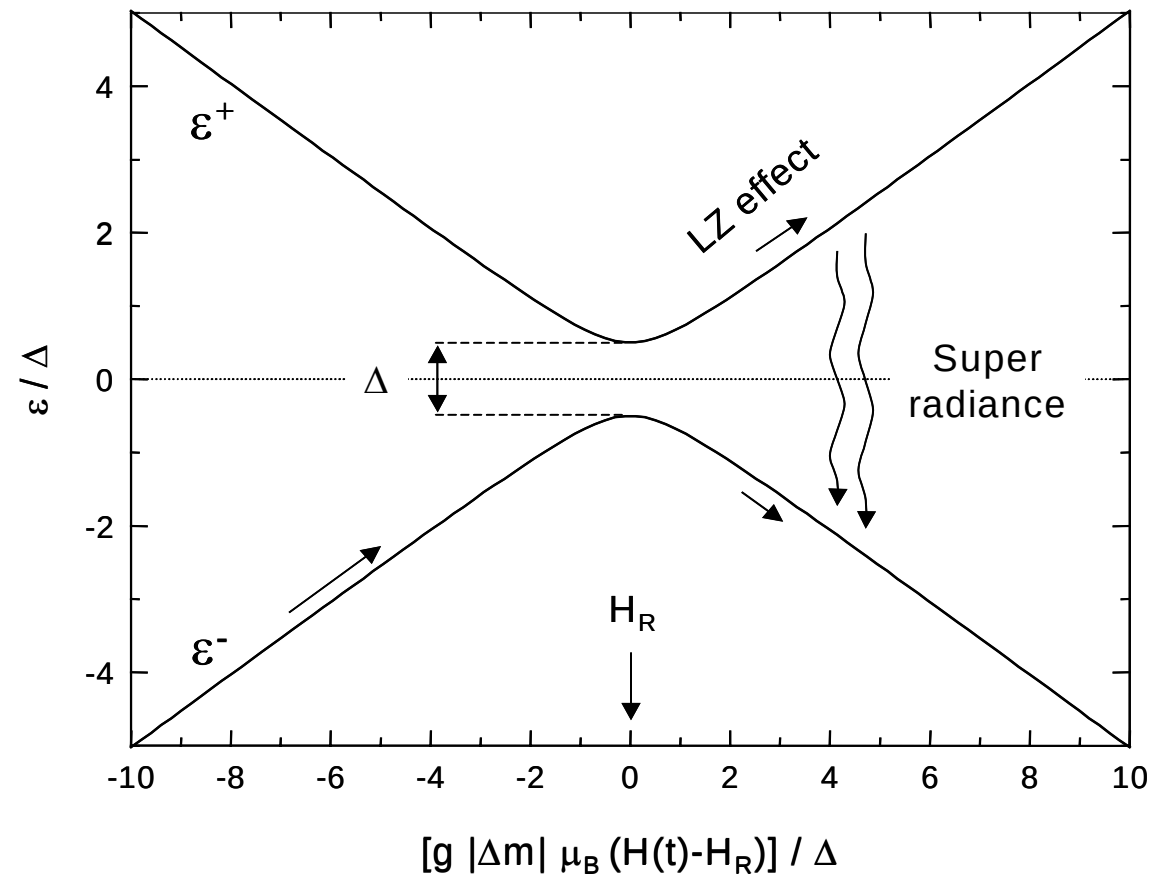

Figure 3. A pair of tunnel-split levels $\varepsilon^+/\varepsilon^-$ versus the energy bias $W = \varepsilon_m-\varepsilon_{m'} = g\ |m-m'|\ \mu_B\ (H(t)-H_R)$. $H_R$ denotes the (third) resonance field. The total magnetization reversal occurs after crossing the resonance $H(t) > H_R$ via superradiant magnetic dipolar transitions between the levels m and m', with unperturbed energies $\varepsilon_m$ and $\varepsilon_{m'}$, respectively.

During the second stage, these excited states decay due to the superradiance onto the lower branch $\varepsilon^-$ (Fig. 3). In the limit of a very small sweep rate, $\varepsilon \gg 1$, almost all molecules follow the lower energy branch, so that the evolution of the system is entirely determined by the Landau-Zener effect and the superradiance is irrelevant. At a high sweep rate, $\varepsilon \ll 1$, the majority of the molecules initially cross to the upper branch and then decay to the lower branch due to the superradiance. At this stage $s_z=M_z/M$ (**M** being the magnetic moment of the system) satisfies the following equation [22]:

$$\frac{d}{dt}s_z(t)=\frac{\alpha}{\hbar}\left(1-s_z^2(t)\right)W(t) \qquad (1)$$

where

$$\alpha=\frac{1}{6}N\left\langle S_z\right\rangle^2 g^2\left(\frac{e^2}{\hbar c}\right)\left(\frac{\Delta}{m_e\ c^2}\right)^2 \qquad (2)$$

$N$ is the total number of $Mn_{12}$ molecules in the crystal, $S$=10 is the spin of the molecule, $g$ is the gyromagnetic ratio, $e$ and $m_e$ are the electron charge and mass, $c$ is the speed of light, and $\left\langle S_z\right\rangle=(m'-m)/2$. In the last expression, $m$=-10 and $m'$=7 are the magnetic quantum numbers or the resonant levels at the third resonant field, $\mu_0 H \sim 1.3$ T. The exact solution of Eq. (1) depends strongly on the initial condition for the superradiance stage. The latter is difficult to predict because of the contribution of both coherent and incoherent processes to the initial Landau-Zener stage [23]. However, one observation immediately follows form Eq. (1). Consider crossing of the third resonance, where $H=H_R$, by a linear field sweep, $\delta H=H(t)-H_R= r\, t$. The relation between the energy sweep rate, $v$, introduced earlier, and the field sweep rate, $r$, is $v = 2\, g\, \mu_B \left\langle S_z \right\rangle r$. According to Eq. (1), the dependence of $\sqrt{r}\left(dM_z/d\delta H\right)$ on $\delta H/\sqrt{r}$ must be independent of $r$ if the initial condition for $s_z$ at the beginning of the superradiant relaxation is independent of $r$. Notice that after the coherent Landau-Zener stage, $s_z=1-2P_{LZ}=-1+2\varepsilon$, that is, the initial condition for Eq. (1) does depend on r. However, the solution of Eq. (1) with this initial condition gives:

$$s_z(t) = \tanh\left[\frac{\alpha\, g\, \mu_B \left\langle S_z \right\rangle}{\hbar}\left(\frac{\delta H}{\sqrt{r}}\right)^2 - \frac{1}{2}\ln\frac{1}{\varepsilon}\right] \qquad (3)$$

which has only a logarithmic deviation on $r$ from the proposed scaling.

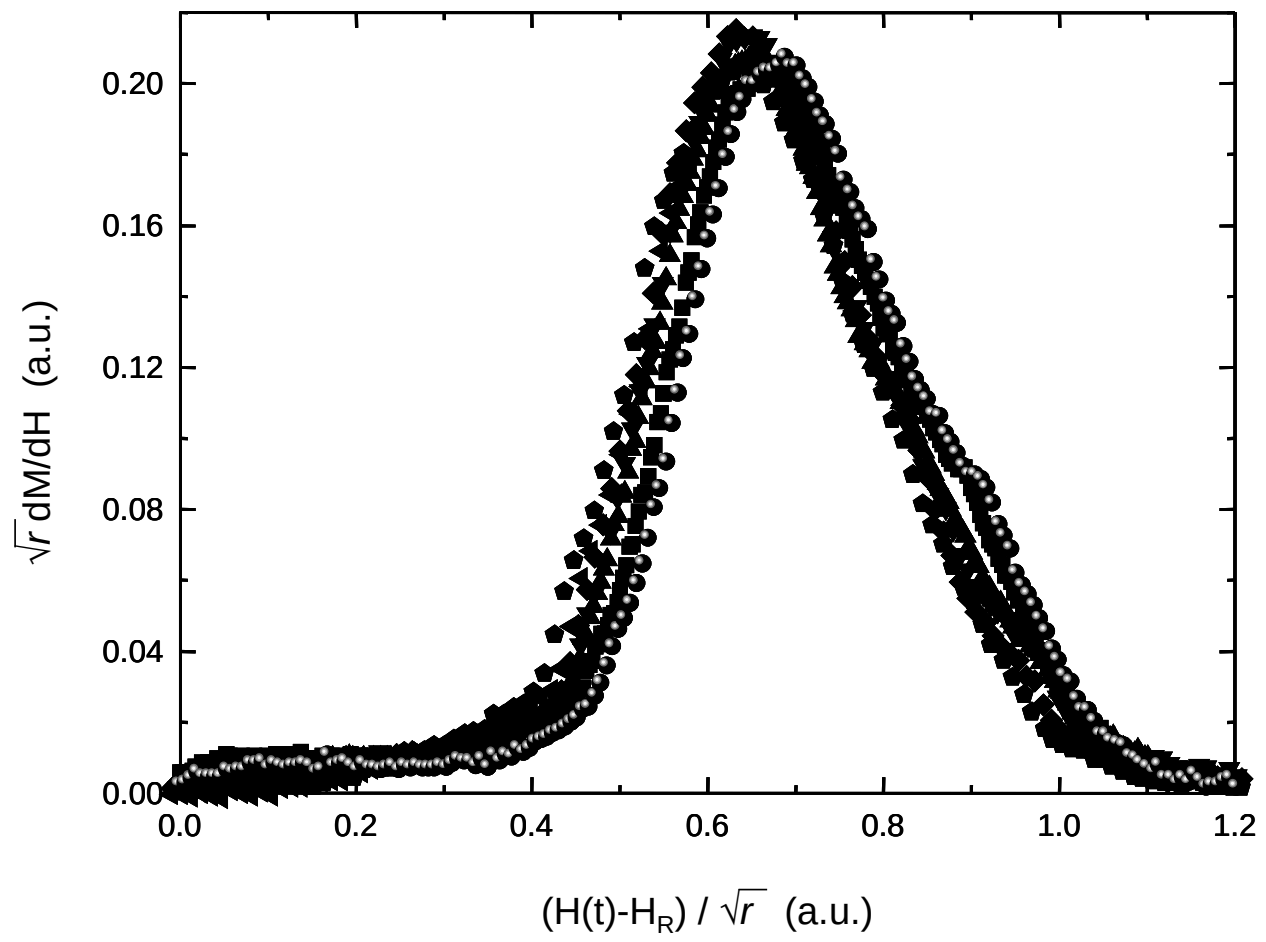

Figure 4. Plot of $\sqrt{r}\left(dM_z/d\delta H\right)$ versus $\delta H/\sqrt{r}$ with $\delta H$=H(t)-$H_R$ at $\mu_0$ $H_R$=1.34 T for all curves at different sweep rates from 1kT/s to 3kT/s.

As can be seen from the inset of Fig. 1, the field sweep in the field range shown in Fig. 2 is with good accuracy linear on time. The scaling of the experimental data along the lines of the above-mentioned theory is shown in Fig. 4. Given the approximations involved, the scaling appears to be rather good. It allows one to estimate the constant $\alpha$ in Eq. (1), $\alpha \sim 10^{-8}$. For $N\sim10^{18}$ this requires $\Delta\sim10^{-3}$ K, which seems to be 3-4 orders of magnitude higher than expected at the third resonance. One should note, however, that the tunnel splitting depends exponentially on the magnetic anisotropy that, in its turn, depends strongly on the elastic stress. It is not inconceivable, therefore, that the magnetostriction effects resulting from a short field pulse are responsible for the high value of the tunnel splitting at the third resonance.

In conclusion, we have found a new spin relaxation effect in a single crystal of $Mn_{12}$ molecular magnets at a high field-sweep rate. The observed dependence of the differential susceptibility on the magnetic field correlates with the theory of collective electromagnetic relaxation.

The Belgian IUAP, the Flemish GOA/2004/02 and FWO have supported this work. J.V. is a postdoctoral fellow of the FWO – Vlaanderen. The work of the group of Barcelona has been supported by the EC through contract No IST-2001-33186 and by Spanish Government through contract No MAT-2002-03144. The work of E.M.C. has been supported by the NSF Grant No. EIA-0310517.